\documentclass[usenatbib]{mn2e}

\usepackage{txfonts}
\usepackage{graphicx}
\usepackage{natbib}

\def\nat{Nature}
\def\apj{ApJ}
\def\mnras{MNRAS}

\def\aap{A\&A}                   
\def\aapr{A\&A Rev}          
\def\apjs{ApJS}                  
\def\apjl{ApJ}                   

\def\aj{AJ}
\def\Msun{M$_{\odot}$}

\begin{document}

\author[E.~Kuiper et al.]{E.~Kuiper,$^1$\thanks{E--mail: kuiper@strw.leidenuniv.nl} B.~P.~Venemans,$^2$ N.~A.~Hatch,$^{3}$ G.~K.~Miley,$^1$ H.~J.~A.~R\"{o}ttgering$^1$ \\
$^1$ Leiden Observatory, Leiden University, P.O. 9513, Leiden 2300 RA, the Netherlands \\
$^2$ European Southern Observatory, Karl--Schwarzschild Strasse, 85748 Garching bei M\"{unchen}, Germany \\
$^3$ School of Physics and Astronomy, The University of Nottingham, University Park, Nottingham NG7 2RD, UK \\
}

\title[Spectroscopy of z$\sim$3 protocluster candidates]{A $z\sim3$ radio galaxy and its protocluster: evidence for a superstructure?}

\maketitle

\begin{abstract}
We present spectroscopic follow-up observations of Lyman Break Galaxies (LBGs) selected in the field surrounding the radio galaxy MRC~0316-257 at $z\sim3.13$ (0316). Robust spectroscopic redshifts are determined for 20 out of 24 objects. Three of the spectroscopically confirmed galaxies have $3.12<z<3.13$ indicating that these objects reside in a protocluster structure previously found around the radio galaxy. An additional 5 objects are found 1600~km~s$^{-1}$ blue-shifted with respect to the main protocluster structure. This is in addition to three [O{\sc iii}] emitters found at this redshift in a previous study. This is further evidence that a structure exists directly in front of the 0316 protocluster. We estimate that the foreground structure is responsible for half of the surface overdensity of LBGs found in the field as a whole. The foreground structure is associated with a strong surface density peak 1.4~Mpc to the North-West of the radio galaxy and a 2D Kolmogorov-Smirnov test indicates that the spatial distributions of the 0316 and foreground galaxies differ at the 3$\sigma$ level. In addition, we compare the properties of protocluster, foreground structure and field galaxies, but we find no significant differences. In terms of the nature of the two structures, a merger scenario is a possible option. Simple merger dynamics indicates that the observed relative velocity of 1600~km~s$^{-1}$ can be reproduced if the two structures have masses of $\sim5\times10^{14}$~\Msun~and have starting separations of around 2.5 to 3~Mpc. It is also possible that the foreground structure is unrelated to the 0316 protocluster in which case the two structures will not interact before $z=0$.
\end{abstract}


\section{Introduction} \label{sec:intro}

The exact role the environment plays in the evolution of galaxies has been a long standing question in astronomy. It has been widely observed that the properties of galaxies depend on environment. High density environments in the local Universe, such as galaxy groups and clusters, are dominated by red early-type galaxies \citep{dressler1980,butcher1984}. This is in contrast with the lower density environments where blue star-forming late-type galaxies are more frequent. Furthermore, elliptical galaxies in cluster environments are generally older than their field counterparts \citep{clemens2006,sanchez2006,vandokkum2007,gobat2008} and the red sequence extends down to fainter magnitudes in denser environments indicating it is formed earlier there \citep{tanaka2005,tanaka2007,tanaka2008}. In addition, cD galaxies, the most massive galaxies known, are located exclusively in galaxy clusters. 

In order to adequately explain the differences between low- and high-density environments it is essential to study galaxy clusters at all epochs. By doing so it may be possible to identify what exact physical processes constitute the more general term of 'environmental influence'. Unfortunately, the search for galaxy clusters at $>1.5$ is difficult and only a handful of spectroscopically confirmed galaxy clusters at $z>1.5$ with X-ray emission are currently known \citep{wilson2008,papovich2010,tanaka2010,henry2010,gobat2011}.

One of the most successful methods to push the search for galaxy clusters beyond $z=2$ is targeting high-z radio galaxies \citep[hereafter HzRGs,][]{miley2008}. With large observed $K$ band luminosities, these galaxies are thought to have large stellar masses of the order of $10^{11}-10^{12}$~\Msun~\citep{roccavolmerange2004,seymour2007}. In the model of hierarchical galaxy formation these massive galaxies should  be located in dense environments and are thus possible members of galaxy cluster progenitors. These structures are often referred to as 'protoclusters', because at these redshifts galaxy clusters are likely still in the process of formation and therefore have not yet virialised \citep[e.g.][]{kuiper2011a}.

By targeting a HzRG field with a narrowband filter chosen such that it contains a strong emission line at the redshift of the radio galaxy, it is possible to select galaxies in a narrow redshift interval around the radio galaxy. This has resulted in evidence that HzRGs indeed probe overdense regions in the early Universe \citep[e.g.][]{pascarelle1996,knopp1997,pentericci2000,kurk2004a,kurk2004b,venemans2007,matsuda2011,kuiper2011b}. 

Although this method is efficient in locating overdensities, it is limited in the number and type of galaxies that can be selected. The emission line most commonly used for these searches is Ly$\alpha$, thus only galaxies with strong Ly$\alpha$ emission are selected. As a consequence, a large number of galaxies that reside in the overdensity are missed altogether. This method is therefore not suitable for in-depth studies that attempt to obtain a complete picture of the protocluster. 

There is a variety of methods that are aimed at selecting high-$z$ galaxies. The most well-known uses the Lyman break to select UV-bright star-forming galaxies (Lyman Break Galaxies or LBGs) in a relatively broad redshift range compared to the narrowband technique. This method was pioneered by \citet{steidel1995} and has been succesfully used in many studies since then. The Lyman break selection technique selects a much larger sample of star forming galaxies than methods relying on Ly$\alpha$ narrowband data, as only 20 per cent of LBGs at a given luminosity also qualify as Ly$\alpha$ emitters taking into account current selection criteria \citep{steidel2000,steidel2011}.

The Lyman break method has been used only sparingly on HzRG fields \citep{intema2006,overzier2008}. This is mainly because the redshift range probed by the LBG criterion is $\sim0.3-0.7$, significantly larger than the redshift range spanned by a typical protocluster ($\Delta z\sim0.03$ or $\Delta v\sim2000$~km~s$^{-1}$ at $z\sim3$). One of the HzRG fields for which the LBG selection criterion has been used is MRC~0316-257 at $z=3.13$. This is one of the best studied HzRGs at $z>2.5$ and it has been shown to host an overdensity of Ly$\alpha$ emitters \citep[][hereafter V05]{venemans2005}. A study by \citet[][hereafter M08]{maschietto2008} has found tentative evidence for a similar overdensity of [O{\sc iii}] emitters. In an attempt to obtain a complete galaxy census of the overdensity, \citet[][hereafter K10]{kuiper2010} assembled photometry of the field in 18 bands ranging from $U$ band to {\it Spitzer} 8~$\mu$m. K10 selected LBGs in the 40~\arcmin$^2$ field around the radio galaxy and detected a small surface overdensity. The most massive and actively star-forming LBGs were found to be located near to the radio galaxy indicating the presence of environmental influence. However, the inability to distinguish protocluster galaxies from field galaxies is likely to diminish any real trend in the data.

In this work we present spectroscopic follow-up of the sample of LBGs composed in K10. By spectroscopically confirming the redshifts of the individual LBG candidates we can determine which galaxies are truly in the protocluster and which are in the field. This will therefore give us a better estimate of the volume overdensity. Also, it allows us to compare the properties of field and protocluster LBGs in a fully self-consistent manner. This is particularly important in determining whether the protocluster environment influences the evolution of its constituent galaxies.

The paper is structured as follows: a brief summary of the sample selection of K10, a description of the data and its reduction are given in Sect.~\ref{sec:data}. Spectroscopic redshifts and the resulting velocity distribution are discussed in Sect.~\ref{sec:results} and further discussion concerning the presence of a possible superstructure is presented in Sect.~\ref{sec:disc}. Finally, conclusions and a future outlook are presented in Sect.~\ref{sec:conc}. Throughout this paper a standard $\Lambda$ cold dark matter ($\Lambda$CDM) cosmology is used, with $H_{\rm 0}=71$~km~s$^{-1}$, $\Omega_{\rm M}=0.27$ and $\Omega_{\Lambda}=0.73$. All magnitudes are given in the AB magnitude system.

\section{Sample selection \& data} \label{sec:data}

In K10 a $UVR$ colour criterion was introduced that was designed to select star-forming galaxies in the redshift range $3.0<z<3.3$.
\begin{eqnarray} \label{eq:crit}
& U-V \ge 1.9, & \nonumber \\
& V-R \le 0.51,  & \\
& U-V \ge 5.07\times (V -R)+2.43, & \nonumber \\
& R \le 26. & \nonumber
\end{eqnarray}
A total of 52 galaxies were found to satisfy the criterion. Photometric redshifts were derived for this sample using the {\sc eazy} code \citep{brammer2008} and broadband photometry in 18 bands ranging from $U$ band to {\it Spitzer} 8~$\mu$m. The initial sample was then reduced to 48 by applying a photometric redshift cut of $2.8<z_{\rm phot}<3.5$.

K10 also constructed an additional sample of 55 potential LBGs (pLBGs). These objects satisfy all selection criteria, except that they are too blue to make the $U-V \ge 1.9$ cut. However, all these objects are undetected in the $U$ band used and deeper $U$ band data may yield redder $U-V$ colours. These objects are thus not strictly LBGs when considering the selection criterion of K10, but deeper data may show that they do in fact satisfy all criteria. 

Only objects with $R<25.5$ were considered for follow-up spectroscopy, because the continuum and absorption lines of fainter objects are unlikely to be detected. This reduced the samples to 29 LBGs and 27 pLBGs, respectively. The samples were subsequently divided in three brightness categories: objects with $R<24.5$ are classified as `bright', objects with $24.5<R<25.0$ as `intermediate' and objects with $R>25.0$ as `faint'. To ensure the most detections, the objects in the mask were prioritised according to their brightness. Further restrictions were imposed by the locations of the individual objects as slits in the mask are not allowed to overlap. The final mask contained 13 LBGs and 11 pLBGs of which 10 are classified as `bright', 9 as 'intermediate' and 5 as `faint'. Therefore a total of 24 protocluster candidates were observed spectroscopically. One of these objects is the Ly$\alpha$ emitter \#1867 from V05. This galaxy has been spectroscopically confirmed to be at the redshift of the protocluster.

The spectroscopy was performed with the FOcal Reducer and low dispersion Spectrograph (FORS2) in the mask multi-object spectroscopy mode (MXU) at the Very Large Telescope during the nights of 10 and 11 December 2010. The seeing varied during the two nights between 0.7\arcsec and 1.2\arcsec. The width of the slits in the mask was 1.0\arcsec. The objects were observed through the ``300V'' grism and GG435 blocking filter, with a resolution of 440. The spectral range covered is approximately $4500<\lambda<8500$~\AA. The pixels were binned $2\times2$, which resulted in a spatial scale of 0.25\arcsec~pixel$^{-1}$ and a dispersion of 3.36~\AA~pixel$^{-1}$. A total of 25 exposures of 1560 seconds each were obtained. Between the individual exposures, the pointing of the telescope was shifted in steps of 0.25\arcsec~along the slit to enable more accurate sky subtraction and cosmic ray removal. The total integration time per object was 39,000 seconds (10.83 hr).

Data reduction was performed with various {\sc iraf}\footnote[1]{{\sc iraf} is distributed by the National Optical Astronomy Observatory, which is operated by the Association of Universities for Research in Astronomy, Inc., under cooperative agreement with the National Science Foundation.} routines. The reduction included the following steps: individual frames were bias subtracted and flat fielded using lamp flats. Cosmic rays were identified and removed before the background was subtracted. The background subtracted two-dimensional frames were combined and one-dimensional spectra were extracted. Wavelength calibration was performed using arc lamp spectra and night sky lines in the science frames. The uncertainty in the wavelength calibration is $\sim0.3$~\AA, which corresponds to a systematic redshift uncertainty of $\sigma_{z}\sim0.0002$.

\section{Results} \label{sec:results}

\subsection{Redshift determination}

Spectroscopic redshifts are obtained for 20 objects out of the 24 observed. For the other 4 objects no continuum or emission lines are detected. Examining the data we find artefacts in two of the 2D spectra indicating slit defects as the possible cause of these two non-detections. The remaining two undetected objects either have a low surface brightness or are very faint ($R\sim25.5$) making it impossible to obtain a spectroscopic redshift. Properties of the objects for which a spectrum was extracted are given in Table~\ref{table1}.

The spectra of the objects that do allow for a redshift determination are shown in Fig.~\ref{fig:spec}. The spectra have been plotted in the restframe to facilitate comparison between the different objects. Also, the locations of the most important spectral features have been marked. The 2D spectra have also been included because the Ly$\alpha$ break is more obviously apparent in the 2D spectra.

The results obtained from the spectra are summarised in Table~\ref{table1b}. Spectroscopic redshifts based on both emission and absorption lines are listed as outflows may affect the spectroscopic redshift determined from Ly$\alpha$ \citep[e.g.][]{shapley2003}. For the spectroscopic redshift we take the mean value of the redshifts obtained for the individual discernible absorption features. Uncertainties listed in Table~\ref{table1} are calculated by varying the spectra according to a normal distribution characterised by the rms noise level. The individual lines are subsequently fitted again. This process is repeated 1000 times for each of the spectral features.

Approximately half of the objects show an emission line, which is assumed to be Ly$\alpha$. This is consistent with other spectroscopic studies of LBGs at $z\sim3$ \citep[e.g.][]{shapley2003,steidel2011}. The emission line is used for determining a preliminary redshift. Based on this redshift the spectrum is searched for consistent absorption lines. For objects that do not show an emission line the redshift is determined by identifying multiple interstellar absorption lines such as Si{\sc ii}$\lambda1260$, C{\sc ii}$\lambda1335$ or C{\sc iv}$\lambda1549$ in combination with a possible spectral break. Almost all objects show either a combination of an emission and an absorption line or multiple absorption lines, resulting in robust redshifts. 

There is one object where the redshift is possibly ambiguous. Object \#12 has only one identifiable absorption feature and evidence for a break. Since the Lyman break is a poor redshift indicator, this makes it difficult to set an accurate redshift. The single absorption feature is very strong and based on the approximate location of the break, the line can be identified as either the O{\sc i}/Si{\sc ii} doublet at $\sim1303$~\AA~or C{\sc ii} at 1335~\AA. This indicates its redshift is either $z\sim3.11$ or $z\sim3.01$. To ascertain which is the more likely, the 2D spectrum of \#12 is compared to the 2D spectra of other objects with strong breaks and clear O{\sc i}/Si{\sc ii} and C{\sc ii} features. The former option better resembles the other 2D spectra and therefore we conclude that the redshift of \#12 is $z=3.1127$.

\begin{figure*}
\resizebox{\hsize}{!}{\includegraphics{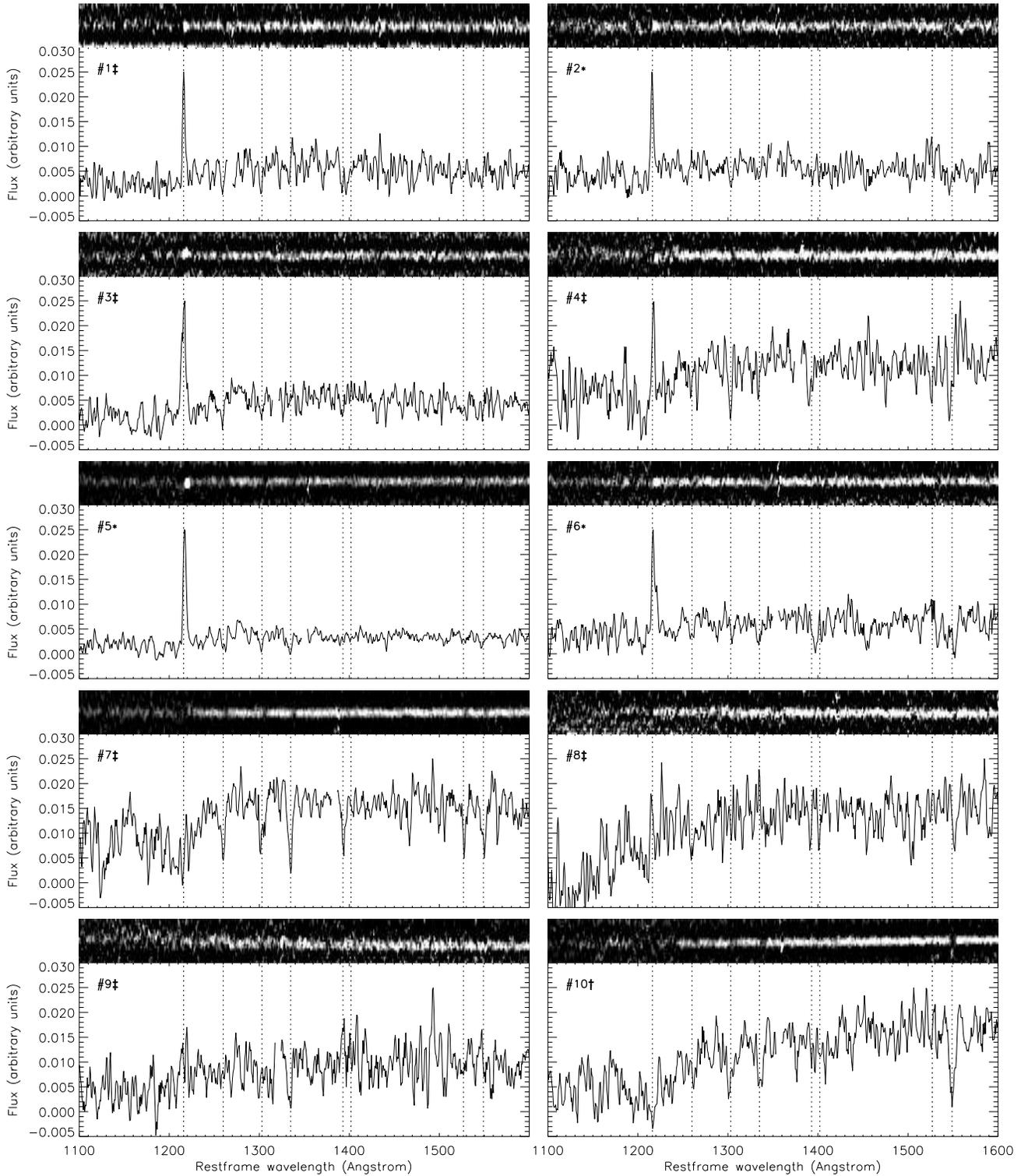}}
\caption{\label{fig:spec} 1D and 2D restframe spectra of individual objects for which a redshift can be determined. The 1D spectra have been scaled to a common arbitrary flux scale. Vertical dotted lines denote the most important spectral features in this wavelength range. From left to right these features are: Ly$\alpha$, N{\sc v}$\lambda1240$, Si{\sc ii}$\lambda1260$, O{\sc i}/Si{\sc ii}$\lambda1303$, C{\sc ii}$\lambda1335$, Si{\sc iv}$\lambda1392,1402$, Si{\sc ii}$\lambda1527$ and C{\sc iv}$\lambda1549$.  The symbols next to the ID number indicate whether the object is located in the 0316 structure ($\ast$), the foreground structure ($\dagger$) or the field ($\ddagger$).}
\end{figure*}

\begin{figure*}
\resizebox{\hsize}{!}{\includegraphics{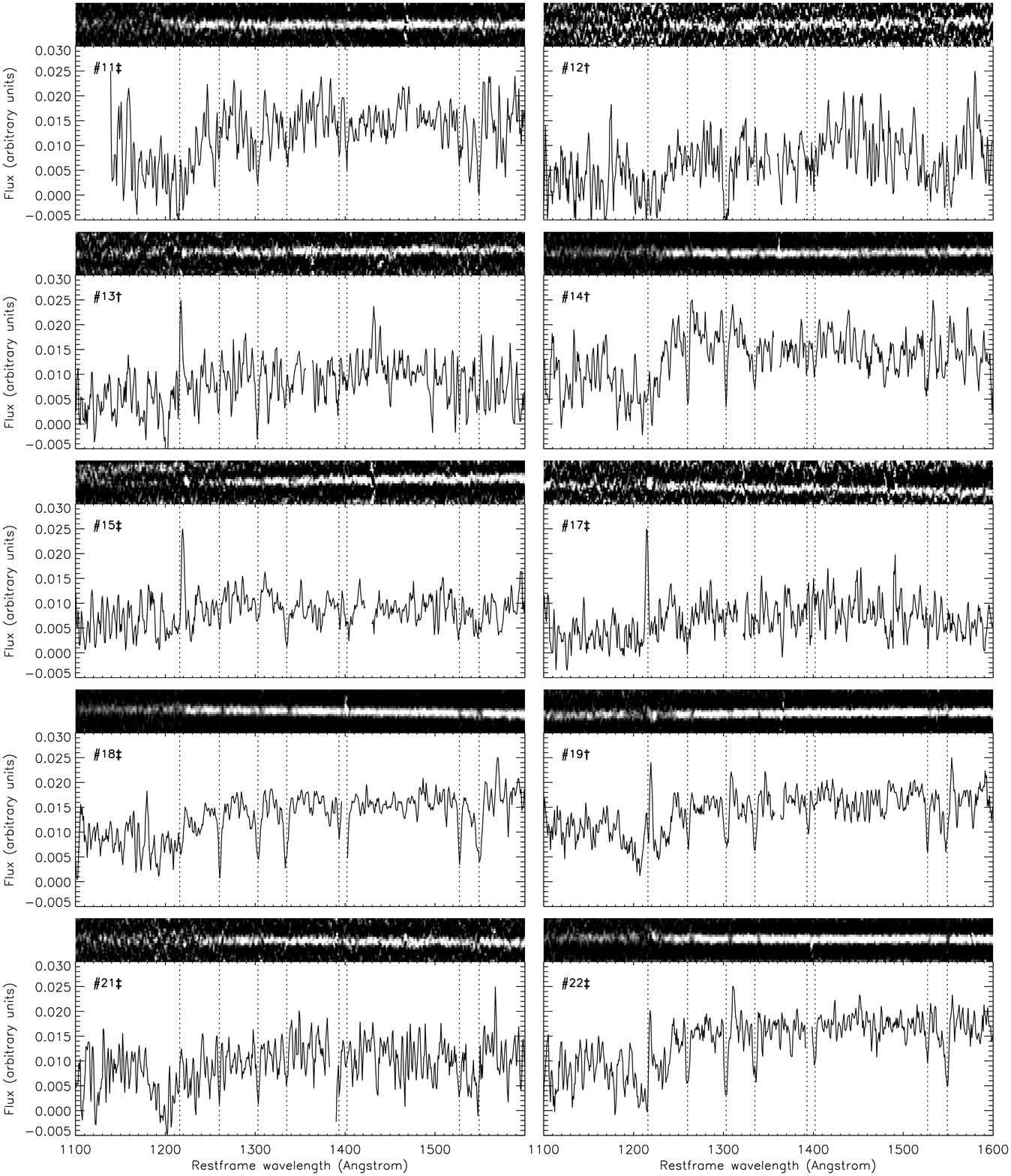}}
\contcaption{\label{fig:spec2}}
\end{figure*}

\begin{table}
\caption{\label{table1} Properties of all LBGs and pLBGs for which spectra can be extracted.}
\small
\begin{tabular}{c|c|c|c|c}
\hline
Object ID & RA & Dec. & Type & $R$ \\
\hline
\#1 & 03:18:01.10 & -25:35:56.1 & pLBG & $25.09$   \\
\#2 & 03:18:12.30 & -25:35:42.2 & pLBG & $24.80$ \\
\#3 & 03:18:08.72 & -25:35:22.4 & LBG & $24.11$ \\
\#4 & 03:18:11.53 & -25:35:08.2 & LBG & $24.52$ \\
\#5 & 03:18:08.94 & -25:34:59.6 & LAE/LBG & $23.77$ \\
\#6 & 03:18:05.35 & -25:34:40.8 & LBG & $24.95$  \\
\#7 & 03:18:07.75 & -25:34:26.1 & pLBG & $24.37$ \\
\#8 & 03:18:18.40 & -25:34:17.1 & LBG & $24.54$\\
\#9 & 03:18:20.28 & -25:34:02.0 & LBG & $25.06$ \\
\#10 & 03:17:58.79 & -25:33:49.9 & pLBG & $24.12$ \\
\#11 & 03:18:08.39 & -25:33:39.6 & LBG & $24.69$ \\
\#12 & 03:17:58.93 & -25:33:27.7 & pLBG & $25.19$  \\
\#13 & 03:18:04.10 & -25:33:09.1 & pLBG & $25.42$ \\
\#14 & 03:18:07.43 & -25:32:51.2 & pLBG & $24.47$ \\
\#15 & 03:18:19.41 & -25:38:16.9 & pLBG & $24.42$  \\
\#16 & 03:18:13.19 & -25:38:06.1 & pLBG & $24.83$ \\
\#17 & 03:17:58.33 & -25:37:59.0 & pLBG & $24.94$ \\
\#18 & 03:17:59.20 & -25:37:37.9 & LBG & $23.80$  \\
\#19 & 03:18:19.73 & -25:37:26.8 & LBG & $23.76$ \\
\#20 & 03:18:03.91 & -25:37:14.1 & LBG & $25.44$ \\
\#21 & 03:18:02.78 & -25:37:05.5 & pLBG & $24.99$  \\
\#22 & 03:18:13.25 & -25:36:39.7 & LBG & $23.94$  \\
\hline
\end{tabular}
\end{table}

\begin{table*}
\caption{\label{table1b} Redshifts, equivalent width of the Ly$\alpha$ line and discernible spectral features of all LBGs and pLBGs for which spectra can be extracted. $^{a}$ Only given for galaxies that show Ly$\alpha$ in emission.}
\small
\begin{tabular}{c||c|c|c|c}
\hline
Object ID  & $z_{\rm spec,Ly\alpha}$ & $z_{\rm spec,abs}$ & $EW_{\rm 0,Ly\alpha}$ (\AA)$^{a}$ & Spectral features \\
\hline
\#1 & $3.4004^{+0.0003}_{-0.0003}$ & $3.3965^{+0.0011}_{-0.0009}$ & $17.9\pm1.1$ & Break, Ly$\alpha$, O{\sc i}/Si{\sc ii}, Si{\sc iv}   \\
\#2 & $3.1258^{+0.0004}_{-0.0004}$ & $3.1282^{+0.0023}_{-0.0026}$ & $14.9\pm0.6$ & Break, Ly$\alpha$, O{\sc i}/Si{\sc ii} \\
\#3  & $3.2306^{+0.0004}_{-0.0005}$ & $3.2303^{+0.0016}_{-0.0015}$ & $43.9\pm2.8$ & Break, Ly$\alpha$, Si{\sc ii}, O{\sc i}/Si{\sc ii}, Si{\sc iv}\\
\#4 & $3.0521^{+0.0005}_{-0.0005}$ & $3.0442^{+0.0011}_{-0.0010}$ & $5.9\pm0.7$ & Break, Ly$\alpha$, O{\sc i}/Si{\sc ii}\\
\#5 & $3.1343^{+0.0001}_{-0.0001}$ & $3.1266^{+0.0011}_{-0.0007}$ & $39.3\pm1.3$ & Break, Ly$\alpha$, O{\sc i}/Si{\sc ii}, C{\sc ii}, C{\sc iv} \\
\#6 & $3.1251^{+0.0010}_{-0.0010}$ & $3.1219^{+0.0016}_{-0.0015}$ & $17.8\pm1.8$ & Break, Ly$\alpha$, Si{\sc ii}, C{\sc ii}  \\
\#7 & - & $3.0324^{+0.0007}_{-0.0007}$ & - & Break, Si{\sc ii}, O{\sc i}/Si{\sc ii}, C{\sc ii}, C{\sc iv}, Si{\sc ii}, C{\sc iv}, Fe{\sc ii}, Al{\sc ii} \\
\#8 & - & $2.9352^{+0.0017}_{-0.0006}$ & - & Break, Si{\sc ii}, C{\sc iv}\\
\#9 & $3.2257^{+0.0042}_{-0.0040}$ & $3.2181^{+0.0034}_{-0.0019}$ & $12.8\pm3.3$ & Break, Ly$\alpha$, C{\sc ii}\\
\#10 & - & $3.1121^{+0.0010}_{-0.0011}$ & - & Break, O{\sc i}/Si{\sc ii}, C{\sc ii}, C{\sc iv}\\
\#11 & - & $2.7795^{+0.0017}_{-0.0019}$ & - & Break, O{\sc i}/Si{\sc ii}, C{\sc ii} \\
\#12 & - & $3.1127^{+0.0017}_{-0.0019}$ & - & Break, O{\sc i}/Si{\sc ii} \\
\#13 & $3.1032^{+0.0017}_{-0.0012}$ & $3.0988^{+0.0015}_{-0.0016}$  & $9.2\pm2.0$ & Break, Ly$\alpha$, O{\sc i}/Si{\sc ii} \\
\#14  & - & $3.1041^{+0.0009}_{-0.0012}$ & - & Break, Si{\sc ii}, O{\sc i}/Si{\sc ii}, C{\sc ii}\\
\#15 & $2.9215^{+0.0006}_{-0.0005}$ & $2.9109^{+0.0015}_{-0.0015}$ & $7.8\pm0.8$ & Break, Ly$\alpha$, C{\sc ii} \\
\#16 & - & - & - & - \\
\#17 & $3.2252^{+0.0006}_{-0.0006}$ & $3.2295^{+0.0046}_{-0.0057}$ & $12.6\pm1.4$ & Break, Ly$\alpha$, Si{\sc ii}\\
\#18 & - & $2.9865^{+0.0005}_{-0.0007}$ & - & Break, Si{\sc ii}, O{\sc i}/Si{\sc ii}, C{\sc ii}, C{\sc iv}, Fe{\sc ii}, Al{\sc ii} \\
\#19 & $3.1115^{+0.0007}_{-0.0007}$ & $3.1003^{+0.0016}_{-0.0009}$ & $2.5\pm0.5$ & Break, Ly$\alpha$, Si{\sc ii}, O{\sc i}/Si{\sc ii}, C{\sc ii}, C{\sc iv}, Fe{\sc ii}, Al{\sc ii} \\
\#20 & - & - & - & - \\
\#21 & - & $3.0233^{+0.0023}_{-0.0013}$ & - & Break, O{\sc i}/Si{\sc ii}, Si{\sc ii} \\
\#22 & - & $2.9996^{+0.0006}_{-0.0007}$ & - & Break, Si{\sc ii}, O{\sc i}/Si{\sc ii}, C{\sc ii}, C{\sc iv}, Fe{\sc ii}, Al{\sc ii} \\
\hline
\end{tabular}
\end{table*}

\subsection{Redshift distribution} \label{sec:distz}

\begin{figure*}
\resizebox{\hsize}{!}{\includegraphics{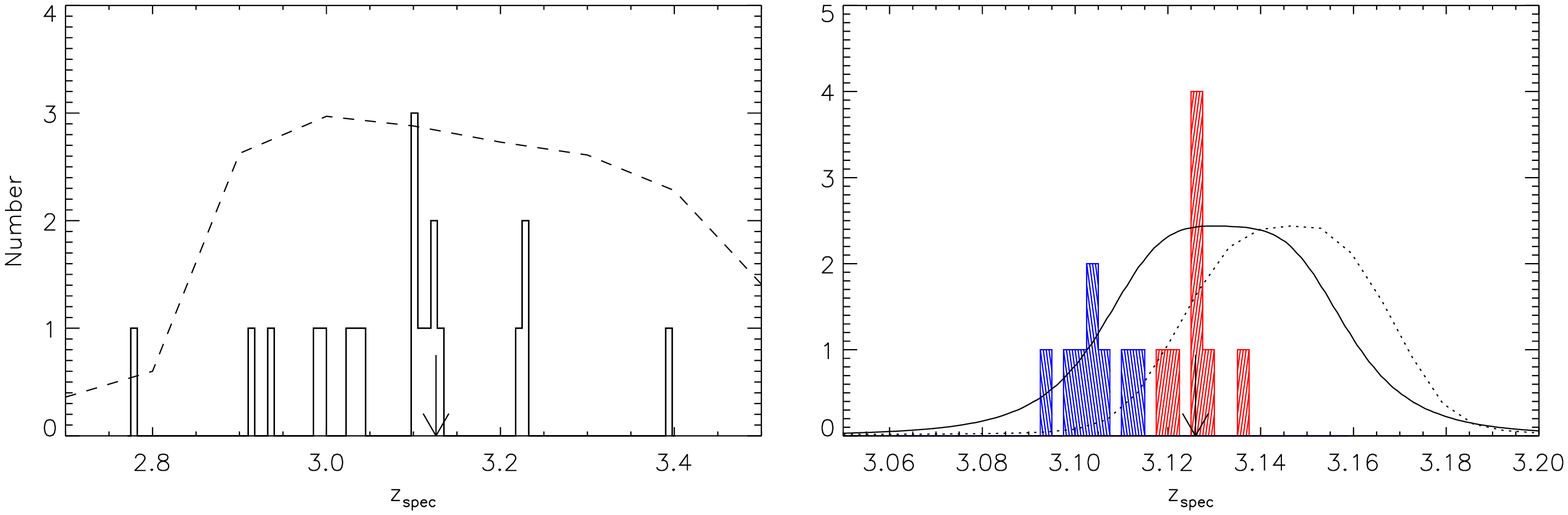}}
\caption{\label{fig:histz} Left panel: Distribution of all spectroscopic redshifts obtained in this work. The dashed curve shows the selection efficiency of the LBG selection criterion. Right panel: Redshift distribution of [O{\sc iii}] emitters and LBGs that are located between $3.08 < z < 3.18$. Also included is LAE \#1518 of V05 which has been identified as an LBG. The objects in the red dashed histogram are those LBGs that are associated with the 0316 radio galaxy, whereas the blue dashed histogram includes [O{\sc iii}] emitters and LBGs that are likely in a foreground structure. The solid and dotted curves indicate the transmission curves of the narrowband filters used to select the [O{\sc iii}] emitters and LAEs, respectively. The arrow marks the redshift of the radio galaxy in both of the panels. All Ly$\alpha$ based redshifts are corrected for the commonly observed shift between Ly$\alpha$ emission and absorption lines.}
\end{figure*}

The full distribution of spectroscopic redshifts based on the absorption lines is shown in the left panel of Fig.~\ref{fig:histz}. The redshift of the radio galaxy is marked by an arrow. All spectroscopically confirmed LBGs have redshifts consistent with $2.7 < z < 3.5$. The low-$z$ interloper rate is therefore small. Assuming that the four non-detected LBG candidates are low-$z$ galaxies the worst-case success rate is $\sim83$ per cent.

The left panel of Fig.~\ref{fig:histz} also shows a clear concentration of galaxies near the redshift of the radio galaxy, consistent with the presence of a structure. To verify this we compare the observed $z_{\rm spec}$ distribution to the selection efficiency curve also shown in Fig.~\ref{fig:histz}. The curve indicates that the LBG selection criterion of K10 is most efficient in selecting objects between $2.8 < z < 3.5$.  A Kolmogorov-Smirnov (KS) test shows that there is a probability of 0.012 that the observed spectroscopic distribution is drawn from the distribution defined by the selection efficiency curve. The two distributions therefore differ at the $2.5\sigma$ level.

As M08 found evidence for a possible foreground structure at $z\sim3.1$, we take a closer look at the redshift interval $3.05<z<3.20$. This is shown in the right panel of Fig.~\ref{fig:histz}. The distribution includes the [O{\sc iii}] emitters of M08 and the relevant LBGs presented in this work. We correct the redshifts of the M08 [O{\sc iii}] emitters that have Ly$\alpha$ based redshifts, because the Ly$\alpha$ line is commonly redshifted with respect to the absorption lines. This redshift is due to outflows and the resonant nature of the Ly$\alpha$ line. The applied correction is taken to be the mean difference in redshift of all galaxies in our sample that show both Ly$\alpha$ emission and absorption lines. This correction is $\Delta z\sim 0.005$ or $\Delta v\sim350$~km~s$^{-1}$, which is roughly consistent with the shift found by \citet{shapley2003}. 

The LAEs of V05 are not included in Fig.~\ref{fig:histz} because the narrowband filter used to select this sample does not allow detection of objects at $z\sim3.1$. This is illustrated by the dotted line in Fig.~\ref{fig:histz}. The distribution of LAEs is thus strongly concentrated near $z\sim3.13$ by design and including it would unfairly skew the overall distribution. The exception to this is LAE \#1518. K10 identified this object as an LBG and as such it is included in the right panel of Fig.~\ref{fig:histz}. It is also included in the subsequent analysis where possible.

The work of M08 aimed to identify [O{\sc iii}] emitters in the 0316 protocluster. Spectroscopic follow-up of three of the candidate [O{\sc iii}] emitters showed that these objects are not located at $z\sim3.13$ as expected but at $z\sim3.1$. Taking into account the correction for the Ly$\alpha$ redshifts, this amounts to a $\sim1700$~km~s$^{-1}$ blueshift with respect to the radio galaxy. The confirmation of an additional 5 objects at $z\sim3.1$ presented in this work further strengthen the notion that a structure exists in front of the 0316 protocluster. 

The question is whether the structures are separate or whether they belong to one larger protocluster. When we look in detail at the LAE distribution of V05 we find that the latter is a possibility. The velocity dispersion of this sample is 535~km~s$^{-1}$ and the median redshift is similar to the median redshift of the red distribution shown in Fig.~\ref{fig:histz}. If we use a KS test to compare this to the expected distribution based on the transmission of the Ly$\alpha$ narrowband filter (dotted line), then we find a probability of $1.2\times10^{-6}$ that the LAE distribution is drawn from the expected distribution. This indicates that the blueshift of the LAE distribution with respect to the central redshift targeted by the narrowband filter is real. This in turn implies that the distribution of LAEs may extend to lower redshifts, but that these objects have been missed due to the location of the narrowband filter.

 On the other hand, a normalized tail index \citep{bird1993} of 0.66 implies that the composite distribution more closely resembles a uniform distribution rather than a single Gaussian. For this work we will follow the approach of M08 and assume that the foreground structure is a separate structure. More evidence for this will be provided in Sect.~\ref{sec:spatial}.
 
Assuming that there are two subgroups we find mean redshifts of $z=3.1039$ and $z=3.1262$ which implies a velocity difference of $\sim1620$~km~s$^{-1}$. Using a Gapper scale estimator \citep{beers1990} we find a velocity dispersion of $965\pm112$~km~s$^{-1}$ for the composite distribution and individual velocity dispersions of $492\pm120$~km~s$^{-1}$ and $364\pm120$~km~s$^{-1}$ for the blue- and redshifted subgroups, respectively.

Although we cannot determine, based on the redshift distribution, whether the two groups are separate structures or one larger protocluster, for the remainder of this work we will refer to the objects associated with the possible foreground structure as foreground objects, whereas those objects at $z\sim3.13$ will be classified as 0316 galaxies. Objects not associated with either of the $z\sim3.1-3.13$ structures will be referred to as field galaxies.

\section{Discussion} \label{sec:disc}

\subsection{A possible superstructure and implications for the overdensity} \label{sec:odens}

In K10, the LBG surface density of the 0316 field was found to be a factor $1.6\pm0.3$ larger than the control field used. Using a volume argument and assuming that the surface overdensity is caused by a single structure connected to the $z=3.13$ radio galaxy, K10 translated this surface density to a volume density which is $8\pm4$ larger than the field density. 

Based on the surface overdensity, one expects spectroscopic follow-up to reveal approximately 1 out of 3 objects to be in the 0316 protocluster. Instead, 3 out of 20 objects are found to be associated with the radio galaxy and an additional 5 objects are found to be located at $\sim3.1$. We therefore confirm the surface overdensity, but it is not solely caused by the 0316 protocluster. The volume overdensity around the 0316 radio galaxy found by K10 is thus not as large as previously estimated and must be adjusted.

To correct the volume overdensity we must assess what fraction of the overdensity is due to the foreground structure.  Due to the small samples that are considered, this cannot be more than a rough estimate. In the sample of 13 [O{\sc iii}] emitters found by M08 a total of eight have spectroscopic redshifts; five due to overlap with the Ly$\alpha$ emitter sample of V05 and three through spectroscopic confirmation of the [O{\sc iii}] line. The latter three were all found to be at $z\sim3.1$. Taking into account that the five unconfirmed [O{\sc iii}] emitters may be in either of the two structures, the fraction of foreground objects is thus 25--60~per~cent. 

In this work, five out of eight objects (or $\sim60$~per~cent) are found to be in the foreground structure. Based on these numbers we assume that half of the surface overdensity can be attributed to the foreground structure. We therefore estimate the volume overdensity of the 0316 protocluster to be $\sim4$, rather than 8. This is very similar to the overdensity of Ly$\alpha$ emitters of $3.3^{+0.5}_{-0.4}$ found in V05. Based on these numbers we also expect that the foreground structure is similar in richness and mass as the 0316 protocluster.

\subsection{Spatial distribution} \label{sec:spatial}

\begin{figure}
\resizebox{\hsize}{!}{\includegraphics{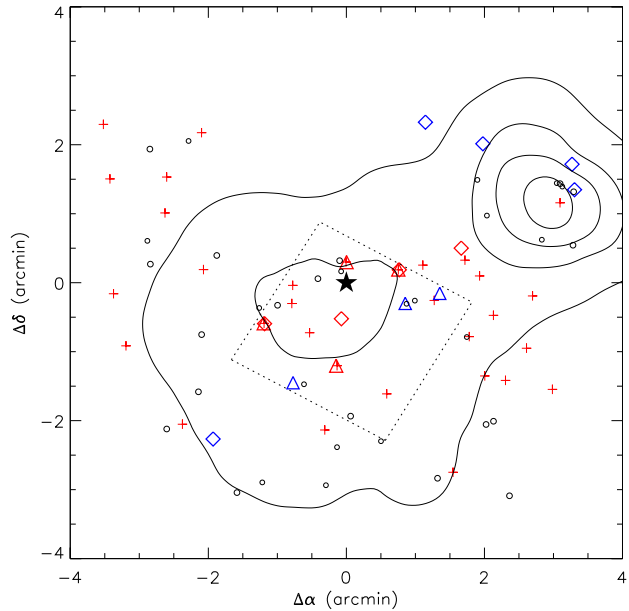}}
\caption{\label{fig:spatdist} Spatial distribution of all spectroscopically confirmed objects that reside in the 0316 protocluster (red symbols) or the foreground structure (blue symbols). Ly$\alpha$ emitters are denoted by plus signs, [O{\sc iii}] emitters by triangles and spectroscopically confirmed LBGs/pLBGs by diamonds. Also shown are the locations of the unconfirmed LBGs and pLBGs as small open circles. The location of the radio galaxy is marked by the star. The dotted lines denote the size and location of the narrowband image used for the detection of the [O{\sc iii}] emitters. The contours indicate the surface density of unconfirmed LBGs and the LBGs that are shown to be either at $z\sim3.10$ or $z\sim3.13$.}
\end{figure}

Figure~\ref{fig:spatdist} shows the spatial distribution of all objects that are spectroscopically confirmed to be either in the 0316 protocluster or in the $z=3.1$ foreground structure. This includes the 32 Ly$\alpha$ emitters of V05 and the three [O{\sc iii}] emitters of M08. The blue objects are those identified to be in the foreground structure. 

Interestingly, four out of five of the foreground LBGs are located in a small area in the North-West region of the field. This specific region is also mostly devoid of $z=3.13$ Ly$\alpha$ emitters. The North-West region is thus dominated by foreground objects. Eight of the non-confirmed LBGs are also located in that general region, four of which are strongly clustered.

The distribution of the foreground [O{\sc iii}] emitters does not reflect this apparent concentration of objects. This is, however, due to the small size of the narrowband image used for the detection of these objects, as illustrated by the outline shown in Fig.~\ref{fig:spatdist}. This limits the location of the foreground [O{\sc iii}] emitters to the central region of the field.

To further illustrate the subclustering, Fig.~\ref{fig:spatdist} also shows LBG surface density contours. The LBGs that have been shown to be field galaxies are not considered, but LBGs that have not been spectroscopically confirmed or have been confirmed to be in either of the two $z\sim3$ structures are included. The contours have been obtained using a grid with a gridsize of 3\arcsec. For each cell of the grid the surface density was calculated by determining the distance to the $N$th nearest neighbour and subsequently using $\sigma=N/\pi r_N^2$. The contours shown have been obtained with $N=8$. The resulting surface density map is smoothed with a smoothing length of 0.5\arcmin. A clear peak in the surface density map is located in the North-West region, near the concentration of foreground objects and $\sim3$\arcmin~or $\sim1.4$~Mpc from the radio galaxy. A second, less pronounced peak in the surface density is located in the centre of the field, near the radio galaxy. This provides further evidence that the foreground structure is offset from the 0316 protocluster.

The significance of the spatial subclustering can be quantified using a 2D KS test. First we determine whether the distribution of LBGs in the foreground structure is consistent with being drawn from a random distribution. We find a probability of 0.033, which implies that the distribution is different from random at the $\sim2\sigma$ level. 

When the spatial distribution of foreground objects is compared to that of the 0316 objects, we find a probability of 0.0034 that both originate from the same parent distribution. The distributions of the two structures therefore differ at the $\sim3\sigma$ significance level. The foreground [O{\sc iii}] emitters have not been taken into account in this comparison. The foreground structure thus seems to be offset from the 0316 structure. We consider this evidence that the two structures are two separate groups of galaxies and not one single protocluster.

\subsection{Influence of environment on galaxy properties} \label{sec:galprop}

The influence of environment on galaxy evolution is an important topic in present day astronomy. Local galaxies in dense environments are generally older, redder and have lower star formation rates (SFRs) than those in less dense environments. There has, however, been mounting evidence that the decrease in star formation observed locally in dense regions turns around at earlier cosmic times \citep{elbaz2007,cooper2008,tran2010,hilton2010,popesso2011}. 

Protocluster fields make excellent targets for studying environmental effects at $z>2$ and several studies have presented ample evidence that environment influences galaxy properties at $z\sim2$. \citet{tanaka2010b} showed that galaxies in the $z\sim2.15$ protocluster around PKS~1138-262 assembled their mass earlier than field galaxies. \citet{hatch2011b} found that H$\alpha$ emitters in the protocluster around the radio galaxy 4C+10.48 at $z=2.35$ are twice as massive as their field counterparts. Similarly, \citet{steidel2005} showed that galaxies in a serendipitously discovered protocluster at $z=2.3$ are approximately twice as old and twice as massive as their field counterparts. 

For the 0316 protocluster, no significant differences have been found between the field and the protocluster galaxies in terms of mass or SFR (K10). There are, however, trends of decreasing mass and SFR with increasing distance from the radio galaxy. One of the main problems is that the field interlopers in the LBG sample of K10 possibly dilute any differences that may be apparent in a pure protocluster sample. With the spectroscopy presented in this work a first division between protocluster LBGs and field LBGs can be made.

\subsubsection{SED fitting}

We first examine the properties of the LBGs in different environments by fitting their spectral energy distributions using photometry presented in K10. For this we use the {\sc fast} SED fitting code \citep{kriek2009} in combination with the \citet{bruzual2003} evolutionary population synthesis models. Originally, all objects were assumed to be located at $z=3.13$. The addition of spectroscopic redshifts allows for a fully self-consistent comparison between field and protocluster galaxies. The free parameters in the fitting routine are the age, mass, SFH and the extinction by dust, but as in K10 we only focus on the stellar mass, because this is the only property that can be determined with reasonable accuracy. 

Briefly summarising the details of the SED fitting process employed in K10: we consider exponentially declining SFHs with decay times, $\tau$, ranging from 10~Myr to 10~Gyr with steps of 0.1~dex. The ages we consider range from log(age/yr)=7 to the age of the Universe at $z\sim3.13$ which is log(age/yr)=9.3. The \citet{calzetti2000} extinction law is used for the internal dust extinction, with $A_{\rm V}$ ranging from 0 to 3 with steps of 0.1. For all cases a Salpeter mass function and solar metallicity are assumed.

\begin{figure}
\resizebox{\hsize}{!}{\includegraphics{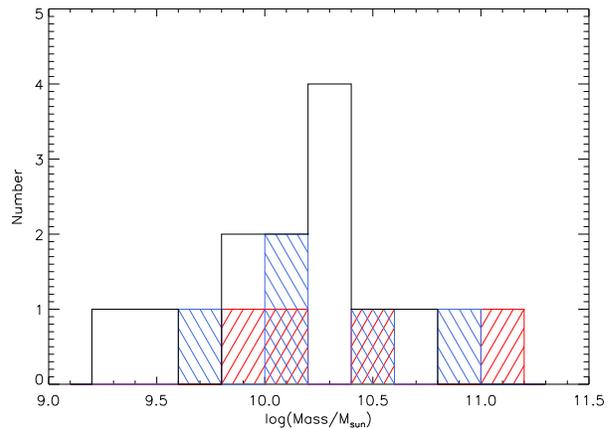}}
\caption{\label{fig:masshisto} Stellar mass distributions of galaxies residing in the field (black), the 0316 protocluster (red dashed) and the foreground structure (blue dashed).}
\end{figure}

The SED fitting results show a marginally larger mean stellar mass of $4.8\times10^{10}$~\Msun~for the 0316 galaxies compared to $2.9\times10^{10}$~\Msun~for the foreground structure and $1.7\times10^{10}$~\Msun~for the field galaxies. When we combine the 0316 and foreground samples we obtain a mean mass of $3.7\times10^{10}$~\Msun~for LBGs in dense environments. However, the small samples considered in this work imply that this difference is not significant. This is also apparent from the stellar mass distributions shown in Fig.~\ref{fig:masshisto}. The small samples make it impossible to distinguish the distributions. KS tests also reflect this, yielding probablities of $0.75-0.9$ that the various distributions are drawn from the same parent distribution. There is therefore no discernible difference in stellar mass between the field and protocluster populations.

\subsubsection{Stacked spectra}

\begin{figure}
\centering
\includegraphics[width=83mm]{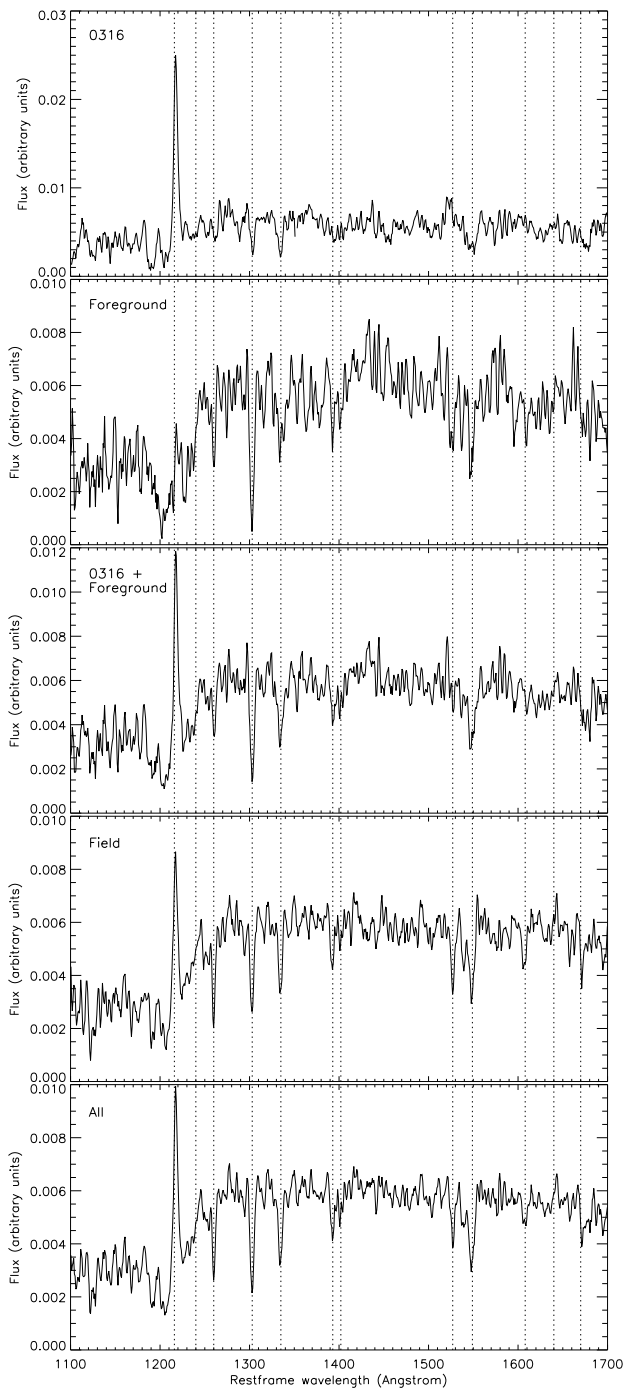}
\caption{\label{fig:stacks} Stacked LBG spectra including, from top to bottom: 0316 objects with $3.12<z<3.13$, foreground structure objects with $3.1<z<3.12$, combined sample of 0316 and foreground objects, field objects and all spectroscopically confirmed objects. Dotted vertical lines indicate the location of the most important spectral features as in Fig.~\ref{fig:spec}. Spectral features above 1600~\AA~are Fe{\sc ii}$\lambda1608$, He{\sc ii}$\lambda1640$ and Al{\sc ii}$\lambda1670$.}
\end{figure}

The presence or absence of environmental dependence is studied further using the stacked spectra. In Fig.~\ref{fig:stacks} we show a series of stacked spectra of the galaxies in each of the environmental categories. These stacked spectra have been obtained by shifting the individual spectra to a common restframe wavelength scale using the absorption line redshifts. The individual spectra are then scaled to the same mean flux level in the restframe wavelength range $1300<\lambda_{\rm rest}<1500$~\AA~and subsequently added together. Here the observed wavelength range $5565<\lambda<5590$~\AA~is excluded due to the presence of strong night-skyline residuals. Since we are dealing with small samples no other outliers are excluded in the stacking process. 

The properties of all detectable spectral lines in the stacked spectra are listed in Table~\ref{table2}. Uncertainties on the derived properties were obtained by repeating the stacking process, but with a number of the spectra replaced by randomly drawn spectra from the same sample. This is to obtain a measure of the intrinsic scatter between the different spectra. For the field galaxies we replace three of the spectra, for the combined 0316+foreground sample we replace two spectra, whereas for the 0316 and foreground structures only one spectrum is replaced. In total 10, 15, 20 and 50 fake spectra are constructed for the 0316, foreground, 0316+foreground and field galaxies, respectively. The fake stacked spectra are subsequently varied according to their rms noise after which all properties are recalculated. For each fake spectrum this process is repeated 100 times. The standard deviations of the subsequent distributions are taken as $1\sigma$ uncertainties. The Ly$\alpha$ FWHM are measured from stacked spectra using the Ly$\alpha$ redshifts rather than the absorption line redshifts. Using the latter increases the FWHM by a factor of 1.5-2.

\begin{table*}
\centering
\caption{\label{table2} Properties of the emission and absorption lines found in the stacked spectra of the 0316 protocluster, the foreground structure and the field. Restframe equivalent widths are taken to be positive for emission lines and negative for absorption lines. Velocity offsets are given with respect to the O{\sc i}/Si{\sc ii} doublet. FWHM values are corrected for the instrumental resolution. $^a$ Based on a stack of the three objects that do show Ly$\alpha$ emission. $^b$ These values are obtained from stacked spectra created using the Ly$\alpha$ redshift where available. $^c$ This value cannot be constrained and is therefore not listed. $^{d}$ The UV slope $\beta$ is calculated using the $R$ and $I$ band data of K10.}
\begin{tabular}{l|c|c|c|c}
\hline
  & 0316 & Foreground & 0316+Foreground & Field \\
\hline
\hline
$\Delta v_{\rm Ly\alpha}$ (km s$^{-1}$) & $+442\pm142$  & $+734\pm225^{a}$ & $+451\pm156$  & $+396\pm162$\\
$EW_{\rm 0,Ly\alpha}$ (\AA) & $26.4\pm3.8$  & $-13.3\pm5.3$ & $11.3\pm3.7$  & $7.4\pm2.1$\\
FWHM$_{\rm Ly\alpha}$ (km s$^{-1}$)$^b$ & $561\pm118$ & -$^c$ & $493\pm122$ & $803\pm241$\\
\hline
$\Delta v_{\rm SiII}$ (km s$^{-1}$)  & - & $+308\pm178$ & $+340\pm167$ & $+281\pm152$   \\
$EW_{\rm 0,SiII}$ (\AA) & - & $-1.9\pm0.8$ & $-1.5\pm0.8$ &  $-2.2\pm0.8$ \\
FWHM$_{\rm SiII}$ (km s$^{-1}$) & - & $400\pm352$ & -$^c$  & $558\pm212$ \\
\hline
$\Delta v_{\rm OI/SiII}$ (km s$^{-1}$)  & $0$ & $0$ & $0$ & $0$    \\
$EW_{\rm 0,OI/SiII}$ (\AA) & $-2.1\pm0.7$ & $-4.5\pm1.1$ & $-3.5\pm0.9$ & $-2.4\pm0.5$ \\
FWHM$_{\rm OI/SiII}$ (km s$^{-1}$) & $365\pm233$ & $799\pm233$ & $708\pm223$  & $623\pm201$\\
\hline
$\Delta v_{\rm CII}$ (km s$^{-1}$)  & $-263\pm169$ & $+103\pm360$ &  $-117\pm202$ & $-61\pm188$   \\
$EW_{\rm 0,CII}$ (\AA) & $-3.0\pm1.1$ & $-3.1\pm1.1$ & $-2.7\pm0.8$ & $-1.8\pm0.6$ \\
FWHM$_{\rm CII}$ (km s$^{-1}$) & $726\pm309$ & $1802\pm545$ & $1127\pm559$ & $509\pm205$\\
\hline
$\Delta v_{\rm SiIV}$ (km s$^{-1}$)  & - & - & $+151\pm196$ & $-61\pm204$   \\
$EW_{\rm 0,SiIV}$ (\AA) & - & - & $-1.4\pm0.8$  & $-0.9\pm0.5$\\
FWHM$_{\rm SiIV}$ (km s$^{-1}$) & - & - & -$^c$  & $102\pm226$ \\
\hline
$\Delta v_{\rm SiII}$ (km s$^{-1}$)  & - & $-75\pm263$ & - & $+88\pm216$    \\
$EW_{\rm 0,SiII}$ (\AA) & - & $-2.1\pm0.9$ & - & $-1.7\pm0.5$ \\
FWHM$_{\rm SiII}$ (km s$^{-1}$) & - & $711\pm266$ & - & $315\pm231$ \\
\hline
$\Delta v_{\rm CIV}$ (km s$^{-1}$)  & $-369\pm267$ & $-194\pm220$ & $-213\pm226$ & $-120\pm324$   \\
$EW_{\rm 0,CIV}$ (\AA) & $-5.3\pm1.5$ & $-3.1\pm0.9$ & $-3.4\pm1.1$  & $-2.7\pm0.8$ \\
FWHM$_{\rm CIV}$ (km s$^{-1}$) & $1881\pm414$ & $800\pm334$ & $1408\pm607$ & $887\pm286$ \\
\hline
$\Delta v_{\rm FeII}$ (km s$^{-1}$)  & - & - & -  & $+24\pm252$   \\
$EW_{\rm 0,FeII}$ (\AA) & - & - & - & $-1.2\pm0.5$\\
FWHM$_{\rm FeII}$ (km s$^{-1}$) & - & - & - & $414\pm278$ \\
\hline
$\beta^{d}$ & $-1.7\pm0.2$& $-0.8\pm0.4$ & $-1.2\pm0.3$ & $-1.3\pm0.1$ \\
\hline
\end{tabular}
\end{table*}

The top panel of Fig.~\ref{fig:stacks} shows the stacked spectrum of the three objects identified to be in the 0316 protocluster. The main feature of the spectrum is the strong Ly$\alpha$ emission with $EW_{\rm 0}=26.4\pm3.8$~\AA, but this is mostly driven by the LAE included in the sample. Removing this galaxy lowers the $EW_{\rm 0}$ to 17.9~\AA. The field galaxies, in the fourth panel, also show Ly$\alpha$ emission, but with an $EW_{\rm 0}=7.4\pm2.1$~\AA~it is on average not as strong as in the 0316 protocluster. 

The average spectrum of the foreground galaxies shows no Ly$\alpha$ emission, but considering the small sample size we are unable to determine whether this is a significant difference. In fact, \citet{shapley2003} show that approximately half of all LBGs show Ly$\alpha$ in emission and the other half shows Ly$\alpha$ in absorption. With sample sizes of 3 for the 0316 protocluster and 5 for the foreground structure, the chance that the observed difference between the 0316 and foreground structure is a mere statistical fluctuation should be considered significant. We can therefore draw no conclusions based on this sample, but It would be interesting to see whether this difference persists in a larger sample.

We can slightly alleviate the problem with the sample size by combining the 0316 and foreground structure samples and comparing this to the field population. Any difference would be a strong indication of environmental influence at $z\sim3$. The stacked spectrum is shown in the third panel of Fig.~\ref{fig:stacks} and the relevant properties are listed in Table~\ref{table2}. The spectrum shows Ly$\alpha$ emission with $EW_{\rm 0}=11.3\pm3.7$~\AA. This is consistent with the field population within the $1\sigma$ uncertainties. The strength of the absorption lines of the combined sample are also consistent with that found for the field population. The composite spectrum of galaxies in the $z\sim3$ overdense structures therefore does not differ from field galaxies, so we find no evidence for environmental influence at $z\sim3$.

The properties listed in Table~\ref{table2} can be compared to the results of \citet[][hereafter S03]{shapley2003}. In S03 the spectra of $\sim1000$ LBGs were stacked to perform a detailed study of the average properties of these galaxies. In general, the LBG properties in Table~\ref{table2} are similar to those in \citet{shapley2003}. There is a velocity difference of $\sim400-900$~km~s$^{-1}$ between the Ly$\alpha$ line and the absorption lines, where the Ly$\alpha$ line is redshifted with respect to the absorption. This is also seen in the individual LBG spectra of S03 and is indicative of outflows. The LBGs in the 0316 field therefore also show evidence of outflows.

\citet{shapley2003} divided their sample of LBGs into four bins based on Ly$\alpha$ equivalent width, ranging from Ly$\alpha$ in absorption to strong Ly$\alpha$ emission. LBGs with strong Ly$\alpha$ emission were found to have weaker low-ionisation lines, bluer UV slopes and smaller kinematic offsets between Ly$\alpha$ and interstellar absorption lines. Based on the stacked spectra presented here we can make a similar division between the various samples. For this purpose we will only consider the field population and the combined 0316+foreground population as the individual 0316 and foreground samples are too small to make a meaningful comparison. Both populations fall into the moderate emission category or group 3 of S03.

Little difference is seen between the composite field spectrum and the results of S03. All $EW_0$ values are consistent within 1$\sigma$ with the properties of group 3 in S03. Since we make no distinction in $EW_{\rm Ly\alpha}$ when stacking the spectra we also expect this spectrum to match closely to the full LBG stack of S03. Indeed, all absorption line equivalent widths are fully consistent with the average LBG of S03.

When comparing the 0316+foreground objects with group 3 of S03 we see that the O{\sc i}/Si{\sc ii} doublet and the C{\sc ii} line of the combined sample is stronger. This is partially due to the inclusion of \#12 in the stack, which has an exceptionally strong O{\sc i}/Si{\sc ii} doublet. Removing this object from the summed spectrum reduces the equivalent width to $-2.7\pm0.6$~\AA~which is formally consistent with the results of S03. This, however, does not explain the strong C{\sc ii} feature. The other spectral features are consistent with S03. 

Following the physical picture presented by S03, stronger absorption lines may be explained by a larger covering fraction of the outflowing gas. This would, however, also diminish the Ly$\alpha$ flux. Since the 0316+foreground sample does show significant Ly$\alpha$ emission there must be something compensating for the larger covering fraction. This could be related to a lower than expected dust content. The UV slopes of galaxies are sensitive to dust content, but the values listed in~\ref{table2} show no significant difference between this work and S03. Furthermore, we must consider that the samples used in this study are much smaller than the samples presented in S03. This could indicate that the strong C{\sc ii} absorption line is due to statistical fluctuation. It is therefore necessary to increase the number of spectroscopically confirmed galaxies in order to put proper constraints on any possible differences between the various populations in this work and the results of S03.

The stacked spectra can also be used to determine whether IGM absorption blueward of the Ly$\alpha$ line is more prevalent in either of the structures or in the field. To do this we assess the mean flux level for $\lambda_{\rm rest}<1185$~\AA~and compare it to the mean flux level for $\lambda_{\rm rest}>1280$~\AA. The ratio of these flux levels is highest for 0316 at $0.67\pm0.02$, whereas the foreground and field galaxies show ratios of $0.53\pm0.02$ and $0.50\pm0.01$, respectively. The combined sample of the two structures yields $0.59\pm0.01$. The Ly$\alpha$ break is thus less pronounced in the protocluster galaxies indicating that there is less IGM absorption in the overdense structures.

\subsection{Interacting or unrelated structures?}

The presence of a foreground structure is not the first indication that HzRG-selected protoclusters are part of superstructures. \citet{kuiper2011a} has shown that the well-studied protocluster around PKS~1138-262 (1138) at $z=2.15$ exhibits a broad bimodal velocity structure which has been independently found for both the megaparsec scale structure and the central kiloparsec scale structure. This is best explained by a line-of-sight merger scenario of two massive halos. Could this also be the case for the 0316 protocluster and its foreground companion?

In Sect.~\ref{sec:spatial} we show that the spatial distributions of the two structures in the 0316 field are not drawn from the same distribution and Fig.~\ref{fig:spatdist} indicates a projected separation of $\sim1.4$~Mpc. If there is a merger it is not along the line of sight.

The velocity difference between the 0316 protocluster and the foreground structure is $\sim1600$~km~s$^{-1}$, but this is only the line-of-sight velocity component. If this is indeed a merging or interacting system, then additional transverse velocity components may be present. This implies that the true relative velocity may be larger than 1600~km~s$^{-1}$. The relative velocity is thus similar to the 1600~km~s$^{-1}$ found in the 1138 system. \citet{kuiper2011a} showed that the Millennium simulation \citep{springel2005,delucia2007} could reproduce such a velocity difference, but only for the largest halo masses. Doing the same analysis at $z\sim3$ as in \citet{kuiper2011a} reveals no such mergers in the Millennium simulation.

In order to determine whether such a merger is possible at $z\sim3$, we calculate how the relative velocity evolves with decreasing distance $d$ in the case of two merging massive halos. For this we use the equations described in \citet{sarazin2002}. Conservation of energy dictates
\begin{equation}
\frac{1}{2}m v^{2}-\frac{GM_{\rm 1}M_{\rm 2}}{d}=-\frac{GM_{\rm 1}M_{\rm 2}}{d_{\rm 0}}
\end{equation}
with $m=M_{1}M_{2}/(M_{1}+M_{2})$ and $d_{\rm 0}$ the separation between the structures when they drop out of the Hubble flow. For simplicity it is assumed that the transverse velocity is zero. This yields
\begin{equation}
v=\sqrt{2G(M_{1}+M_{2})\left(\frac{1}{d}-\frac{1}{d_{0}}\right)}.
\end{equation}

\begin{figure}
\resizebox{\hsize}{!}{\includegraphics{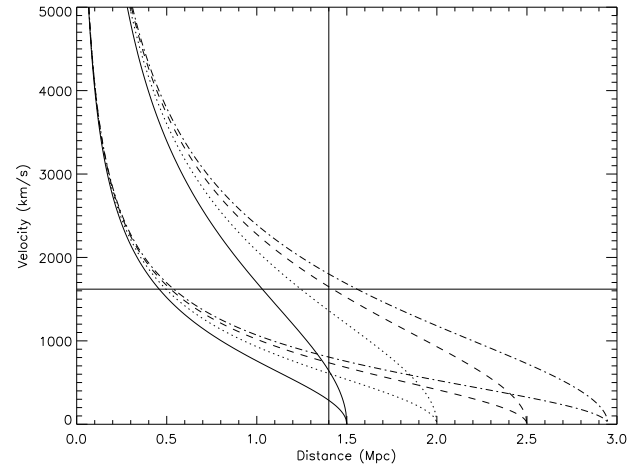}}
\caption{\label{fig:mergvelos} Evolution of the relative velocity of two massive structures with distance for a variety of starting distances $d_0$ and halo masses. The starting separations are 1.5, 2.0, 2.5 and 2.95~Mpc. These are denoted by the solid, dotted, dashed and dash-dotted curves respectively. For each starting distance two halo masses are considered. The halo mass ratio is fixed at 1:1 and the massed considered are 1 and $5\times10^{14}$~\Msun~with the steeper curves corresponding to the larger halo masses. The horizontal and vertical lines indicate the relative velocity of 1600~km~s$^{-1}$ and the projected separation between the two structures as observed in the 0316 field.}
\end{figure}

In Fig.~\ref{fig:mergvelos} we show how the velocity increases with decreasing distance in a merger scenario. The curves shown are for a variety of values for $d_{\rm 0}$. The largest value for $d_0$ was chosen such that the time it takes to reach a relative velocity of 1600~km~s$^{-1}$ is equal to the age of the Universe at $z=3.13$. We also consider two specific halo masses, but in all cases it is assumed that the 0316 and foreground structure are of equal mass. We see that for masses of $10^{14}$~\Msun~a velocity of $\sim1600$~km~s$^{-1}$ is only reached at small separations of the order of $<0.6$~Mpc. For masses closer to the estimated mass of the 0316 protocluster, the distance at which $1600$~km~s$^{-1}$ is reached ranges between $1.0 < d < 1.6$~Mpc.

As we have shown in Sect.~\ref{sec:spatial}, the projected distance between the structures is of the order of $\sim1.4$~Mpc. If the two structures are interacting then the true distance is likely larger. The relative velocity of 1600~km~s$^{-1}$ is also likely a lower limit due to projection effects. If the merger scenario is possible we therefore expect the curves to cross through the upper right quadrant of Fig.~\ref{fig:mergvelos}. There are two curves that meet this requirement, with starting separations of 2.5 and 2.95~Mpc and masses of $5\times10^{14}$~\Msun. It is therefore possible that the system is undergoing a merger, but only if the two structures are massive and the starting separation is roughly 2.5 to 3~Mpc.

The other option is that the 0316 protocluster and the foreground structure are `unrelated' structures in the Hubble flow at the time of observing. Assuming that the Hubble flow dominates the relative motions of the two structures we use the Hubble law to find that a velocity difference along the line of sight of 1620~km~s$^{-1}$ implies a distance of at least 23~comoving~Mpc. Following \citet{bahcall2004} we assume that the mean distance between two clusters in the Hubble flow at $z\sim3$ is $\sim50$~Mpc. What is the chance of encountering such a line-of-sight alignment as possibly witnessed here? In a field of $100\times100$~Mpc$^2$ we expect a total of $\sim15$ structures. The 0316 field covers approximately 9~Mpc${^2}$. So the probability of having one additional structure directly in front of the main structure is $\sim1$~per~cent. If we lower the mean distance to $\sim30$~Mpc, then the chance increases to $\sim4$~per~cent. These probabilities are small, but they are not negligible. It is thus not unreasonable to find an unrelated foreground structure and the possibility that this is a change alignment of two unrelated structures in the Hubble flow cannot be rejected.

If the two structures are a chance alignment then they are separated by $\sim23$~Mpc. This does not preclude that at some later epoch the two structures will interact. We try to determine whether two massive haloes separated by $\sim23$~Mpc at $z\sim3$ will merge before $z=0$ by using the Millennium simulation. Only the most massive haloes at $z\sim3$ are considered, because the estimated mass of the 0316 protocluster is large ($\sim5\times10^{14}$~\Msun, V05) and the foreground structure seems to have a similar mass. Using a lower limit of $2\times10^{13}$~\Msun~we find a total of 20 halo pairs with relative distances between 20 and 25~Mpc. For the majority of the pairs the separation decreases, but none reach separations smaller than 8~Mpc. None of the pairs therefore interact or merge before $z=0$. Thus, if the two structures in the 0316 field are not interacting at $z\sim3$, then it is likely that there will be no interaction between the structures on the timescale of $\sim10$~Gyr.

\section{Conclusions} \label{sec:conc}

We have presented spectroscopic follow-up of LBGs in the field surrounding the 0136 protocluster at $z\sim3.13$. We observed a total of 24 LBG candidates. Using spectroscopic redshifts we distinguish between field and protocluster galaxies. This in turn allows us to make a self-consistent comparison between the field and protocluster galaxy samples.

\begin{enumerate}

\item{
We determine redshifts for 20 out of 24 objects, finding that all objects are located between $2.7 < z < 3.5$. This implies an interloper fraction of at most $\sim17$~per~cent. Out of the 20 confirmed objects, 5 are located at $z\sim3.10$ and 3 at $z\sim3.13$. The number of 0316 protocluster objects is too small to account for the surface overdensity presented in K10, but is consistent with the presence of two structures: the 0316 protocluster at $z\sim3.13$ and a foreground structure at $z\sim3.1$. The presence of such a foreground structure was already hypothesised by \citet{maschietto2008}. Based on the redshift distribution, however, it is not clear whether the two structures should be considered as separate or as part of one larger structure.
}

\item{
The spatial distribution of the foreground and 0316 LBGs shows two distinct density peaks: one centred on the radio galaxy and a stronger peak located in the North-West corner of the field. This latter stronger peak coincides with a concentration of foreground objects indicating that the foreground structure is not directly in front of the 0316 protocluster. A 2D Kolmogorov-Smirnov test confirms this, indicating that the spatial distributions of the 0316 and foreground LBGs differ at the $3\sigma$ level. This implies that the two structures are likely indeed separated and not part of a larger protocluster.
}

\item{
The presence of the foreground structure implies that the volume overdensity of LBGs of K10 is overestimated. Instead of the previously determined overdensity of 8, the volume density is only a factor $\sim4$ larger than the field. We also estimate that the foreground structure is of similar mass and richness to the 0316 protocluster.
}

\item{ 
There is no systematic difference in mass between the protocluster galaxies and field galaxies. Stacking the spectra shows that the galaxies associated with the 0316 protocluster have stronger Ly$\alpha$ emission than the field galaxies, whereas the galaxies in the foreground structure show very little Ly$\alpha$ emission. However, considering the limited sample size this may be due to statistical fluctuation. Combining the galaxies in the two structures in one composite sample shows that galaxies in dense environments do not differ from field galaxies. This implies that there is no discernible evidence for environmental effects on galaxy evolution at $z\sim3$. However, the Lyman break is less pronounced in the combined 0316+foreground sample indicating that there is less absorption by the IGM in these structures with respect to the field galaxies.
}

\item{ 
The relative velocity of the 0316 and foreground structures can be reproduced if the structures are merging and have large masses of $5\times10^{14}$~\Msun. Alternatively, the structures may not be merging in which case the relative velocity translates into a radial separation 23~Mpc. Such a large separation at $z\sim3$ means it is unlikely that the two structures will interact by the present day.
}

\end{enumerate}

The results presented here for the different samples of galaxies should be considered preliminary. Further spectroscopic observations are necessary to get a better census of which galaxies are in which of the structures. This will also give better constraints on the potential differences between the protocluster and field galaxies at $z\sim3$. Furthermore, extra data will provide stronger constraints on the merger scenario considered in this work. If additional spectroscopic redshifts result in smaller separations on the sky and in redshift space, then this may increase the likelihood of the merger scenario. 

\section*{Acknowledgements}

We would like to thank the anonymous referee for the very useful comments that have helped improve this paper. This research is based on observations carried out at the European Southern Observatory, Paranal, Chile, with program number 086.A-0930(A). The authors wish to thank the staff at the VLT for their excellent support during the observations. EK acknowledges funding from Netherlands Organization for Scientific Research (NWO). NAH acknowledges support from STFC and the University of Nottingham Anne McLaren Fellowship.


\end{document}